%% file: arXiv_D0Kspiev_20241214.tex
\RequirePackage[mathlines]{lineno} % Display line numbers
\documentclass[a4paper,11pt]{article}
\pdfoutput=1 
\usepackage{jheppub} 
\usepackage[T1]{fontenc} 

\usepackage{threeparttable}%This package is used for tablenotes
\usepackage{multirow}
\usepackage{subfigure}
\usepackage{float}
\let\oldequation\equation
\let\oldendequation\endequation
\renewenvironment{equation}
 {\linenomathNonumbers\oldequation}
 {\oldendequation\endlinenomath}
\begin{document}
\setrunninglinenumbers

\title{\bf\boldmath Study of the semileptonic decay $D^0\rightarrow \bar{K}^0\pi^-e^+\nu_e$}
\collaborationImg{\includegraphics[height=30mm, angle=90]{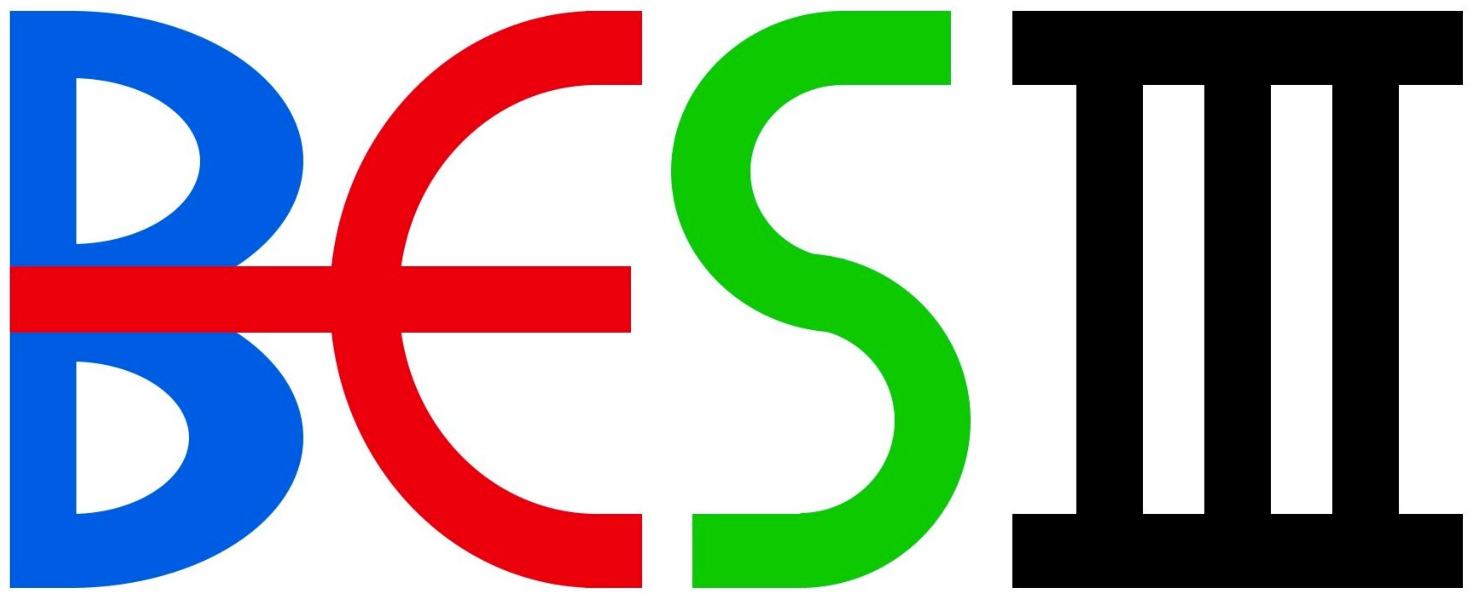}}

\collaboration{The BESIII collaboration}
\emailAdd{BESIII-publications@ihep.ac.cn}

%\begin{linenumbers}

\abstract{
We report an improved study of the semileptonic decay $D^0 \rightarrow \bar{K}^0\pi^-e^+\nu_{e}$ based on 
a sample of $7.9~\mathrm{fb}^{-1}$ of $e^+e^-$ annihilation data  collected at a center-of-mass energy of 3.773~GeV with the BESIII detector at the BEPCII collider. The branching fraction of this decay is measured to be $\mathcal{B}(D^0\rightarrow \bar{K}^0\pi^-e^+\nu_{e}) = (1.444 \pm 0.022_{\rm stat} \pm 0.024_{\rm syst})\%$, which is the most precise to date, where the first uncertainty is statistical and the second is systematic. Based on investigation of the decay dynamics, we find that the decay is dominated by the $K^{*}(892)^-$ component and present an improved measurement of its branching fraction to be $\mathcal{B}(D^0\rightarrow K^{*}(892)^-e^+\nu_e) = (2.039 \pm 0.032_{\rm stat} \pm 0.034_{\rm syst})\%$.
We also determine the ratios of the hadronic form factors for the $K^{*}(892)^-e^+\nu_e$ decay to be $r_{V} = V(0)/A_1(0) = 1.48 \pm 0.05_{\rm stat} \pm 0.02_{\rm syst}$ and $r_{2} = A_2(0)/A_1(0) = 0.70 \pm 0.04_{\rm stat} \pm 0.02_{\rm syst}$, where $V(0)$ is the vector form factor and $A_{1,2}(0)$ are the axial form factors. In addition, the $\bar{K}^0\pi^-$ $\mathcal{S}$-wave component is found to account for $(5.87 \pm 0.32_{\rm stat} \pm 0.16_{\rm syst})\%$ of the total decay rate, corresponding to a branching fraction of $\mathcal{B}[D^0\rightarrow (\bar{K}^0\pi^-)_{S-{\rm wave}}e^+\nu_e] = (0.085 \pm 0.005_{\rm stat} \pm 0.003_{\rm syst})\%$.
}
%\keywords{$\ee$ experiments, charm, leptonic weak decay}
%\arxivnumber{******}

\maketitle
\flushbottom

%%%%%%%%%%%%%%%%%%%%%%%%%%%%%%%%%%%%%%%%%%%%%%%%%%%%%%%%%%%%%%%%
%%%%%     Introduction       Part                  %%%%%%%%%%%%%
%%%%%%%%%%%%%%%%%%%%%%%%%%%%%%%%%%%%%%%%%%%%%%%%%%%%%%%%%%%%%%%%
\section{Introduction}
Studies of semileptonic (SL) decay modes of charm mesons provide valuable information on the weak and strong interactions in mesons composed of heavy quarks~\cite{physrept494}. The SL partial decay width is related to the product of the hadronic form factor describing the strong-interaction in the initial and final hadrons, and the Cabibbo-Kobayashi-Maskawa (CKM) matrix elements $|V_{cs}|$ and $|V_{cd}|$, which parametrize the mixing between the quark flavors in the weak interaction~\cite{prl10_531}. The couplings $|V_{cs}|$ and $|V_{cd}|$ are tightly constrained by the unitarity of the CKM matrix. Hence studies of the dynamics of the SL decays allow measurements of the hadronic form factors, which are important for calibrating theoretical calculations of the relevant strong interaction effects. 

The relative simplicity of theoretical description of the SL decay $D\rightarrow \bar{K}\pi e^+ \nu_e$~\cite{chargeneutral} makes it an optimal place to study the $\bar{K}\pi$ system, and to further determine the hadronic transition form factors.
Measurements of $\bar{K}\pi$ resonant and non-resonant amplitudes in the decay $D^+\rightarrow \bar{K}\pi e^+\nu_e$ have been reported by
the CLEO~\cite{prd74_052001}, BaBar~\cite{prd83_072001} and BESIII~\cite{prd94_032001,prd99_011103} collaborations. In these studies, a nontrivial $S$-wave component is observed along with a dominant $P$-wave.
Furthermore, the form factors in the $D\rightarrow Ve^+\nu_{e}$ transition, where $V$ refers to a vector meson, have
been measured in decays of $D^+\rightarrow \bar{K}^{*0}e^+\nu_e$~\cite{prd74_052001,prd83_072001,prd94_032001}, $D^0\rightarrow K^{*-}e^+\nu_e$~\cite{prd99_011103}, $D^{+,0}\rightarrow \rho^{+,0} e^+\nu_e$~\cite{prl110_131802}
and $D^+\rightarrow \omega e^+\nu_e$~\cite{prd92_071101}, although the precision of these form-factor measurements is limited.

In this paper, an improved measurement of the absolute branching fraction (BF) and the form-factor parameters of the SL decay $D^0\rightarrow \bar{K}^0\pi^-e^+\nu_e$ are reported.
These measurements are performed using an $e^+e^-$ annihilation data sample corresponding to an integrated luminosity of
$7.9~\mathrm{fb}^{-1}$ produced at $\sqrt{s}=3.773$ GeV with the BEPCII collider and collected by the BESIII detector~\cite{Ablikim:2009aa}. 

%%%%%%%%%%%%%%%%%%%%%%%%%%%%%%%%%%%%%%%%%%%%%%%%%%%%%%%%%%%%%%%%
%%%%%     Detector and software Part               %%%%%%%%%%%%%
%%%%%%%%%%%%%%%%%%%%%%%%%%%%%%%%%%%%%%%%%%%%%%%%%%%%%%%%%%%%%%%%
\section{BESIII detector and Monte Carlo simulation}

The BESIII detector~\cite{Ablikim:2009aa} records symmetric $e^+e^-$ collisions provided by the BEPCII storage ring~\cite{Yu:IPAC2016-TUYA01}, which operates with a peak luminosity of $1\times10^{33}$~cm$^{-2}$s$^{-1}$ in the center-of-mass energy range from 1.85 to 4.95~GeV.
BESIII has collected large data samples in this energy region~\cite{Ablikim:2019hff}. The cylindrical core of the BESIII detector covers 93\% of the full solid angle and consists of a helium-based multilayer drift chamber~(MDC), a plastic scintillator time-of-flight system~(TOF), and a CsI(Tl) electromagnetic calorimeter~(EMC), which are all enclosed in a superconducting solenoidal magnet providing a 1.0~T magnetic field. The solenoid is supported by an octagonal flux-return yoke with resistive plate counter muon
identification modules interleaved with steel.
The charged-particle momentum resolution at $1~{\rm GeV}/c$ is
$0.5\%$, and the d$E$/d$x$ resolution is $6\%$ for electrons
from Bhabha scattering. The EMC measures photon energies with a
resolution of $2.5\%$ ($5\%$) at $1$~GeV in the barrel (end cap) region. The time resolution in the TOF barrel region is 68~ps, while that in the end cap region is 110~ps. The end cap TOF system was upgraded in 2015 using multi-gap resistive plate chamber technology, providing a time resolution of 60~ps~\cite{etof}, which benefits 63\% of the data used in this analysis.

Simulated data samples produced with a {\sc geant4}-based~\cite{geant4} Monte Carlo (MC) package,
which includes the geometric description of the BESIII detector and the detector response,
are used to determine detection efficiencies and to estimate backgrounds. The simulation
models the beam energy spread and initial state radiation (ISR) in the $e^+e^-$ annihilations
with the generator {\sc kkmc}~\cite{kkmc}. 
The inclusive MC sample includes the production of $D\bar{D}$ pairs, the non-$D\bar{D}$ decays of the $\psi(3770)$, the ISR
production of the $J/\psi$ and $\psi(3686)$ states, and the continuum processes incorporated in {\sc kkmc}~\cite{kkmc}. 
All particle decays are modelled with {\sc evtgen}~\cite{nima462_152} using branching fractions either taken from
the Particle Data Group~\cite{pdg24}, when available, or otherwise estimated with {\sc lundcharm}~\cite{lundcharm}.
Final state radiation (FSR) from charged final state particles is incorporated using the
{\sc photos} package~\cite{plb303_163}. The generation of signal $D^0\rightarrow \bar{K}^0\pi^-e^+\nu_e$ incorporates knowledge of the form factors  obtained in this work. 

%%%%%%%%%%%%%%%%%%%%%%%%%%%%%%%%%%%%%%%%%%%%%%%%%%%%%%%%%%%%%%%%
%%%%%            Physics Analysis                  %%%%%%%%%%%%%
%%%%%%%%%%%%%%%%%%%%%%%%%%%%%%%%%%%%%%%%%%%%%%%%%%%%%%%%%%%%%%%%
\section{Event selection and data analysis}
The analysis makes use of both ``single-tag'' (ST) and ``double-tag'' (DT) samples of $D$ decays. The ST sample consists of $\bar{D}^0$ decay candidates reconstructed in one of the hadronic final states listed in table~\ref{tab:numST}, which are called the tag decay modes. Within each ST sample, a subset of events is selected where the other tracks in the event are consistent with the decay $D^0\rightarrow \bar{K}^0\pi^-e^+\nu_e$. This subset is referred to as the DT sample. For a specific tag mode $i$, the ST and DT
event yields are expressed as
$$N^{i}_{\rm ST}=2N_{D^0\bar{D}^0}\mathcal{B}^i_{\rm ST}\epsilon^i_{\rm ST},~~
N^{i}_{\rm DT}=2N_{D^0\bar{D}^0}\mathcal{B}^i_{\rm ST}\mathcal{B}_{\rm SL}\epsilon^i_{\rm DT},$$
where $N_{D^0\bar{D}^0}$ is the number of $D^0\bar{D}^0$ pairs, $\mathcal{B}^i_{\rm ST}$ and
$\mathcal{B}_{\rm SL}$ are the BFs of the $\bar{D}^0$ tag decay mode $i$ and the $D^0$ SL decay
mode, $\epsilon^i_{\rm ST}$ is the efficiency for finding the tag candidate, and
$\epsilon^i_{\rm DT}$ is the efficiency for simultaneously finding the tag $\bar{D}^0$ and the SL decay.
The BF for the SL decay is given by
\begin{equation}
  \mathcal{B}_{\rm SL} \,=\,
  \frac{N_{\rm DT}}{\sum_i N^{i}_{\rm ST} \,
      \left(\epsilon^i_{\rm DT}/\epsilon^i_{\rm ST}\right)} \,=\,
  \frac{N_{\rm DT}}{N_{\rm ST} \, \epsilon_{\rm SL}}, \label{eq:branch}
\end{equation}
%where $N_{\rm DT}$ is the total yield of DT events, $N_{\rm ST}$ is the total ST yield, and $\epsilon_{\rm SL} = \left(\sum_i N^{i}_{\rm ST} \, \epsilon^i_{\rm DT}/\epsilon^i_{\rm ST}\right) \,/\, \sum_i N^{i}_{\rm ST}$
where $N_{\rm DT}$ is the total yield of DT events, $N_{\rm ST}$ is the total ST yield, and $\epsilon_{\rm SL}=(\sum_i N^{i}_{\rm ST}\times\epsilon^i_{\rm DT}/\epsilon^i_{\rm ST})/\sum_i N^{i}_{\rm ST}$
is the average efficiency of reconstructing the SL decay in a ST event,
weighted by the measured yields of tag modes in the data.

Charged tracks are required to be well-reconstructed in the MDC detector,
with the polar angle $\theta$ satisfying $|\cos\theta|<0.93$.
Their distances of the closest approach to the interaction point (IP) are required to be less than 10~cm along the beam direction and less than 1~cm in the perpendicular plane.
To discriminate pions from kaons, the $dE/dx$ and TOF information is
combined to obtain particle identification (PID) likelihoods for the pion ($\mathcal{L}_{\pi}$) and
kaon ($\mathcal{L}_K$) hypotheses. Pion and kaon candidates are
selected using $\mathcal{L}_{\pi} > \mathcal{L}_{K}$ and
$\mathcal{L}_{K} > \mathcal{L}_{\pi}$, respectively.

Photon candidates are reconstructed from isolated clusters in the
EMC in the regions $|\cos\theta| \le 0.80$ (barrel) and $0.86 \le
|\cos\theta|\le 0.92$ (end cap). The deposited energy of a cluster
is required to be larger than 25 (50)~MeV in the
barrel (end cap) region, and the opening angle between a shower
and the nearest charged track must be at least $10^\circ$.  
To suppress electronic noise and energy deposits unrelated to the
event, the difference between the EMC time and the event start time
is required to be within [0, 700]~ns. 
To reconstruct a $\pi^0$ candidate via $\pi^0\rightarrow \gamma\gamma$, the invariant mass of the
candidate photon pair must be within $(0.115,~0.150)$~GeV$/c^2$. To improve the momentum resolution, a kinematic fit is performed to constrain the $\gamma\gamma$ invariant mass to the
nominal $\pi^0$ mass~\cite{pdg24}. The $\chi^2$ of this kinematic fit is required to be less than 50. The
fitted $\pi^0$ momentum is used for reconstruction of the $\bar{D}^0$ tag candidates.

The ST $\bar{D}^0$ decays are identified using the beam-constrained mass,
\begin{equation}
M_{\rm BC} = \sqrt{(\sqrt{s}/2)^2-|\vec {p}_{\bar D^0}|^2},
\end{equation}
where $\vec {p}_{\bar D^0}$ is the momentum of the
$\bar{D}^0$ candidate in the rest frame of the initial $e^+e^-$ system. 
To improve the purity of the tag decays, the
energy difference $\Delta E = E_{\bar{D}^0} -\sqrt{s}/2$ for
each candidate is required to be within approximately
$\pm3\sigma_{\Delta E}$ around the fitted $\Delta E$ peak, where
$\sigma_{\Delta E}$ is the $\Delta E$ resolution and
$E_{\bar{D}^0}$ is the reconstructed $\bar{D}^0$
energy in the  initial $e^+e^-$ rest frame. The explicit $\Delta E$ requirements for these three ST modes are
listed in table~\ref{tab:numST}.

\begin{figure}[tp!]
\begin{center}
\includegraphics[width=\linewidth]{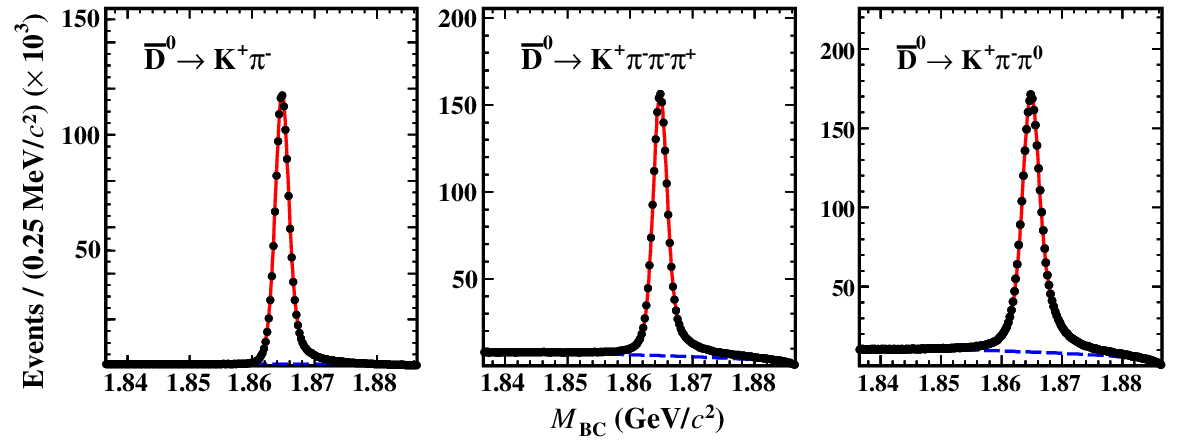}
\caption{(Color online)~The $M_{\rm BC}$ distributions for the three ST modes. The points are data, the solid red curves are the projection of the sum of all fit components and the dashed blue curves are the projection of the background component of the fit.}
\label{fig:tag_md0}
\end{center}
\end{figure}

The distributions of the variable $M_{\rm BC}$ for the three ST
modes are shown in figure~\ref{fig:tag_md0}. Maximum likelihood fits to the $M_{\rm BC}$ distributions are performed. The signal shape is derived from the convolution of the MC-simulated signal function with a double-Gaussian function to account for resolution difference between MC simulation and data. 
An ARGUS function~\cite{plb241_278} is used to describe the combinatorial background shape.
For each tag mode, the ST yield is obtained by integrating the signal
function over the $D^0$ signal region within $1.859<M_{\rm BC}<1.873$~GeV/$c^2$.
In addition to the combinatorial background, there are
also small wrong-sign (WS) peaking backgrounds in the ST $\bar{D}^0$ samples,
from the doubly Cabibbo-suppressed decays of $\bar{D}{}^0\rightarrow K^-\pi^+$, $K^-\pi^+\pi^0$ and $K^-\pi^+\pi^+\pi^-$.
The $\bar{D}{^0}\rightarrow K^0_SK^-\pi^+$, $K^0_S\rightarrow \pi^+\pi^-$ decay shares the same final state as the WS background of $\bar{D}^0\rightarrow K^-\pi^+\pi^+\pi^-$.
The sizes of these WS peaking backgrounds are estimated from simulation, and are subtracted from the corresponding ST yields.
The background-subtracted ST yields and the ST efficiencies for three ST modes are listed in table~\ref{tab:numST}. The total ST yield summed over all three ST modes is 
$N_{\rm ST}=(6306.7\pm2.9)\times 10^3$, where the uncertainty is statistical only.

\begin{table}[tp!]
\caption{ The selection requirements on $\Delta E$, the background-subtracted ST yields, $N_{\rm ST}$, in data, and the ST and DT efficiencies, $\epsilon_{\rm ST}$ and $\epsilon_{\rm DT}$, for each of the three tag decay modes. }
\begin{center}
\begin{tabular}
{lcccc} \hline\hline ST mode      & $\Delta E$ (GeV)     &  $N_{\rm ST}$ ($\times 10^3$)   & $\epsilon_{\rm ST}$  (\%)    & $\epsilon_{\rm DT}$  (\%)   \\
\hline $K^+\pi^-$                & [$-$0.027, 0.027]          &  $1449.3\pm1.3$ & $65.34\pm0.01$     &   $6.69\pm0.01$    \\
       $K^+\pi^-\pi^-\pi^+$    & [$-$0.026, 0.024]          &  $1944.2\pm1.6$ & $40.83\pm0.01$     &   $3.76\pm0.01$   \\
       $K^+\pi^-\pi^0$           & [$-$0.062, 0.049]          &  $2913.2\pm2.0$  & $35.59\pm0.01$    &   $3.54\pm0.01$    \\
\hline\hline
\end{tabular}
\label{tab:numST}
\end{center}
\end{table}

\section{Branching fraction for $D^0\rightarrow \bar{K}^0\pi^-e^+\nu_e$}
Candidates for the SL decay $D^0\rightarrow \bar{K}^0\pi^-e^+\nu_e$ are selected from the remaining tracks recoiling against the ST $\bar{D}^0$ mesons. The $\bar{K}^0$ meson is reconstructed as a $K^0_S \to \pi^+\pi^-$ decay, using two oppositely-charged tracks (with no PID) having an invariant mass within $(0.485,~0.510)$~GeV$/c^2$. For each $K_S^0$ candidate, a fit is applied to constrain the two charged tracks to a common vertex, and this $K^0_S$ decay vertex is required to be separated from the IP by more than twice the standard deviation of the measured flight distance. It is then required that there be only two other tracks with opposite charges in the event.   The track having the same charge as the kaon on the tag side is taken as the electron candidate. 
  For electron PID, the $dE/dx$ and TOF measurements are combined with shower properties from the EMC to construct likelihoods for electron, pion and kaon hypotheses, $\mathcal{L}^{\prime}_e$, $\mathcal{L}^{\prime}_\pi$ and $\mathcal{L}^{\prime}_K$.  The electron candidate must satisfy $\mathcal{L}^{\prime}_{e} > 0.001$ and 
$\mathcal{L}^{\prime}_e/(\mathcal{L}^{\prime}_e+\mathcal{L}^{\prime}_{\pi}+\mathcal{L}^{\prime}_K)>0.8$. Additionally, the EMC energy of the
electron candidate has to be more than 70\% of the track momentum measured in the MDC: $E/p>0.7$.
The energy loss due to bremsstrahlung is partially recovered by adding the energy of the
EMC showers that are within 5$^{\circ}$ of the initial electron direction and not matched to other
particles~\cite{bes3electronSL}.
The final charged track is taken as the pion candidate and must satisfy the same PID criteria as for the ST side. The background from $D^0\rightarrow \bar{K}^0\pi^+\pi^-$ decays reconstructed as $D^0\rightarrow \bar{K}^0\pi^-e^+\nu_e$ is rejected by requiring the $\bar{K}^0\pi^-e^+$ invariant mass ($M_{\bar K^0\pi^-e^+}$) to be less than 1.80~GeV/$c^2$. 
Backgrounds containing additional $\pi^0$ mesons are suppressed by requiring the maximum energy of any unused photon ($E_{\gamma \rm max}$) to be less than 0.25~GeV. 

The energy and momentum carried by the neutrino are denoted by $E_{\rm miss}$ and $\vec{p}_{\rm miss}$, respectively. They are calculated from
the energies and momenta of the tag ($E_{\bar{D}^0}$, $\vec{p}_{\bar{D}^0}$) and the measured SL decay products ($E_{\rm SL}=E_{\bar{K}^0}+E_{\pi^-}+E_{e^+}$,
$\vec{p}_{\rm SL}=\vec{p}_{\bar{K}^0}+\vec{p}_{\pi^-}+\vec{p}_{e^+}$) using the relations $E_{\rm miss}=\sqrt{s}/2-E_{\rm SL}$ and $\vec{p}_{\rm miss}=\vec{p}_{D^0}-\vec{p}_{\rm SL}$   in the initial $e^+e^-$ rest frame. 
Here, the momentum $\vec{p}_{D^0}$ is given by
$\vec{p}_{D^0}=-\hat{p}_{\rm tag}\sqrt{(\sqrt{s}/2)^2-m^2_{\bar{D}^0}},$
where $\hat{p}_{\rm tag}$ is the
momentum direction of the ST $\bar{D}^0$ and $m_{\bar{D}^0}$ is the nominal $\bar{D}^0$ mass~\cite{pdg24}. Information on the undetected neutrino is obtained by using the variable $U_{\rm miss}$ defined by
\begin{equation}
U_{\rm miss} \equiv E_{\rm miss}-|\vec{p}_{\rm miss}|c.
\end{equation}
The $U_{\rm miss}$ distribution
is expected to peak at zero for signal events.

Figure~\ref{fig:formfactor}(a) shows the $U_{\rm miss}$ distribution of the accepted candidates for $D^0\rightarrow \bar{K}^{0}\pi^-e^+\nu_e$ in data.
To obtain the signal yield, an unbinned maximum likelihood fit to the $U_{\rm miss}$ distribution is performed.
In the fit, the signal is described with a shape derived from the simulated signal events convolved with a Gaussian function, where the width of the Gaussian function is determined by the fit. The
background is described by using the shape obtained from the MC simulation.
The fitted yield of DT $D^0\rightarrow \bar{K}^0\pi^-e^+\nu_e$ events is $N_{\rm DT}=8752\pm132$.
The backgrounds from the non-$D^0$ and non-$K_S^0$ decays are estimated by examining
the ST candidates in the $M_{\rm BC}$ and $K^0_S$ sidebands. The yield of this type of background is found to be consistent with zero and is limited to be small enough to not contribute significantly to systematic uncertainties. 
The DT efficiency $\epsilon^i_{\rm DT}$ for each tag mode is summarized in the last column of table~\ref{tab:numST}, and the average DT efficiency $\varepsilon_{\rm SL}$ is estimated to be $(9.79\pm0.01)\%$~\cite{BFK0}.
The difference of the $K^0_S$ reconstruction efficiencies between data and MC is estimated to be $-(1.8\pm0.6)\%$, taking into account the systematic uncertainties due to the tracking efficiencies for the charged pions, and systematic uncertainties associated with the $K^0_S$ mass window and decay length requirements. Hence $\varepsilon_{\rm SL}$ is corrected by $-1.8\%$, 
giving $(9.61\pm0.01)\%$. Hence, the BF obtained is $\mathcal B({D^0\rightarrow \bar{K}^{0}\pi^-e^+\nu_e})=(1.444\pm0.022)\%$. 

Due to the DT technique, the BF measurement is insensitive to the systematic uncertainty in the ST efficiency.
The uncertainty on the electron tracking efficiency (PID) is estimated to be 0.5~(0.1)\% by studying a sample of $e^+e^-\rightarrow \gamma e^+e^-$ events.
The uncertainty due to the pion tracking efficiency (PID) is estimated to be 0.3~(0.3)\% using control samples selected from $D^0\rightarrow K^-\pi^+(\pi^0, \pi^+\pi^-)$ and $D^+\rightarrow K^-\pi^+\pi^+(\pi^0)$. The uncertainty from $K^0_S$ reconstruction is 0.6\%, determined with control samples selected from
$D^0 \rightarrow \bar{K}^0\pi^+\pi^-, \bar{K}^0\pi^+\pi^-\pi^0, \bar{K}^0\pi^0$ and  $D^+ \rightarrow \bar{K}^0\pi^+, \bar{K}^0\pi^+\pi^0, \bar{K}^0\pi^+\pi^+\pi^-$.
 The uncertainty associated with the $E_{\gamma\,{\rm \max}}$ requirement is estimated to be 0.2\% by analyzing DT $D^0\bar{D}^0$ events where $D^0$ mesons decay to hadronic final states of $D^0\rightarrow K^-\pi^+$, $K^-\pi^+\pi^0$ and $K^-\pi^+\pi^+\pi^-$.
The uncertainty due to the modeling of the signal in simulated events is estimated to be 0.7\% by varying the input form-factor parameters determined in this work by $\pm 1\sigma$. The uncertainty associated with the fit of the $U_{\rm miss}$ distribution is estimated to be 1.0\% by varying the fitting ranges and the shapes which parametrize the signal and background, where an asymmetric Gaussian function is used as an alternative signal function. 
The uncertainty associated with the fit of the $M_{\rm BC}$ distributions used to determine $N_{\rm ST}$ is 0.1\% and is evaluated by varying the bin size, fit range and background distributions. Further systematic uncertainties are assigned due to the statistical precision of the simulation, 0.3\%, and the input BF of the decay $K^0_S\rightarrow \pi^+ \pi^-$, 0.1\%. The systematic uncertainty contributions are summed in quadrature, and the total systematic uncertainty on the BF measurement is 1.7\%. Therefore, the BF of $D^0\rightarrow \bar{K}^{0}\pi^-e^+\nu_e$ is determined to be $(1.444\pm0.022_{\rm stat}\pm0.024_{\rm syst})\%$.

%%%%%%%%%%%%%%%%%%%%%%%%%%%%%%%%%%%%%%%%%%%%%%%%%%%%%%%%%%%%%%%%
%%%%%%%%%%    formfactor       Part                %%%%%%%%%%%%%
%%%%%%%%%%%%%%%%%%%%%%%%%%%%%%%%%%%%%%%%%%%%%%%%%%%%%%%%%%%%%%%%
\section{Decay rate formalism for $D^0\rightarrow \bar{K}^0\pi^-$$e^+\nu_{e}$}
The differential decay width of $D^0\rightarrow \bar{K}^{0}\pi^- e^+\nu_{e}$ can be expressed in terms of five kinematic variables:
the squared invariant mass of the $\bar{K}^0\pi^-$ system ($m_{\bar{K}^0\pi^-}^2$), the squared transfer momentum of the $e^+$ and $\nu_e$ ($q^2$), the angle between the $\bar{K}^0$ and the $D^0$ direction in the $\bar{K}^0\pi^-$ rest frame ($\theta_{\bar{K}^0}$), the angle between the $\nu_{e}$ and the $D^0$ direction in the $e^+\nu_e$ rest frame ($\theta_e$), and the acoplanarity angle ($\chi$) between the two decay planes of $\bar{K}^0\pi^-$ and $e^+\nu_e$.
Neglecting the electron mass, the differential decay width of $D^0\rightarrow \bar{K}^{0}\pi^- e^+\nu_{e}$ can be expressed as~\cite{prd46_5040}
\begin{eqnarray}
d^5\Gamma&=&\frac{G^2_F|V_{cs}|^2}{(4\pi)^6m^3_{D^0}}X\beta \mathcal{I}(m_{\bar{K}^0\pi^-}^2, q^2, \theta_{\bar{K}^0}, \theta_e, \chi) dm_{\bar{K}^0\pi^-}^2dq^2d{\rm cos}\theta_{\bar{K}^0}d{\rm cos}\theta_ed\chi,
\label{eq:differential}
\end{eqnarray}
where $X=p_{\bar{K}^{0}\pi^-}m_{D^0}$, $\beta=2p^{*}/m_{\bar{K}^{0}\pi^-}$, $p_{\bar{K}^{0}\pi^-}$ is the momentum of the $\bar{K}^{0}\pi^-$ system in the $D^0$ rest system and $p^*$ is the momentum of $\bar{K}^{0}$ in the $\bar{K}^{0}\pi^-$ rest frame, and $m_{D^0}$ is the known $D^0$ mass~\cite{pdg24}. The Fermi coupling constant is denoted by $G_F$.
The kinematic dependence of the decay density $\mathcal{I}$ is given by
\begin{eqnarray}
\mathcal{I}&=&\mathcal{I}_1+\mathcal{I}_2{\rm cos2}\theta_e+\mathcal{I}_3{\rm sin}^2\theta_e{\rm cos}2\chi+\mathcal{I}_4{\rm sin}2\theta_e{\rm cos}\chi  \nonumber\\
           &+&\mathcal{I}_5{\rm sin}\theta_e{\rm cos}\chi+\mathcal{I}_6{\rm cos}\theta_e+\mathcal{I}_7{\rm sin}\theta_e{\rm sin}\chi \nonumber \\
           &+&\mathcal{I}_8{\rm sin}2\theta_e{\rm sin}\chi+\mathcal{I}_9{\rm sin}^2\theta_e{\rm sin}2\chi,
\label{eq:Ifunc}
\end{eqnarray}
where $\mathcal{I}_{1,...,9}$ depend on $m_{\bar{K}^{0}\pi^-}^2$, $q^2$ and $\theta_{\bar{K}^0}$ and can be expressed in terms of three form factors, $\mathcal{F}_{1,2,3}$~\cite{prd46_5040} . The form factors can be expanded into partial waves including $\mathcal{S}$-wave ($\mathcal{F}_{10}$), $\mathcal{P}$-wave ($\mathcal{F}_{i1}$) and $\mathcal{D}$-wave ($\mathcal{F}_{i2}$), to show their explicit dependences on $\theta_{\bar{K}^0}$.
However, the $\mathcal{D}$-wave component is neglected because of the limited statistics of reconstructed $D^0\rightarrow \bar{K}^{0}\pi^- e^+\nu_{e}$.
Also, no significant $D$-wave contributions were observed for the related decay $D^+\rightarrow K^+\pi^-e^+\nu_e$ in previous BaBar~\cite{prd83_072001} and BESIII~\cite{prd94_032001} analyses.  
% , where the $\mathcal{D}$-wave component are unobserved. 
Consequently, the form factors can be written as
\begin{equation}
\mathcal{F}_1=\mathcal{F}_{10}+\mathcal{F}_{11}\cos\theta_{\bar{K}^0}, \mathcal{F}_2=\frac{1}{\sqrt{2}}\mathcal{F}_{21}, \mathcal{F}_3=\frac{1}{\sqrt{2}}\mathcal{F}_{31},
\label{eq:F1}
\end{equation}
where $\mathcal{F}_{11}$, $\mathcal{F}_{21}$ and $\mathcal{F}_{31}$ are related to the helicity basis form factors $H_{0,\pm}(q^2)$ by~\cite{prd46_5040,RevModPhys67_893}
\begin{eqnarray}
\mathcal{F}_{11} &=& 2\sqrt{2}\alpha q \, H_0 \, \mathcal{A}(m), \nonumber \\
\mathcal{F}_{21} &=& 2\alpha q \, (H_++H_-)   \, \mathcal{A}(m), \nonumber \\
\mathcal{F}_{31} &=& 2\alpha q \, (H_+-H_-)   \, \mathcal{A}(m), 
\label{eq:F2}
\end{eqnarray}
where $\alpha$ is a constant factor, and $\mathcal{A}(m)$ denotes the amplitude of the $\mathcal{P}$-wave component, taken as the Breit-Wigner form given below in Eq.~(\ref{eq:bw}). The helicity form factors can be related to the two axial-vector form factors, $A_1(q^2)$ and $A_2(q^2)$, as well as the vector form factor $V(q^2)$.
The $A_{1,2}(q^2)$ and $V(q^2)$ are all taken as a one-pole form
$A_{1,2}(q^2) = A_{1,2}(0)/(1-q^2/M^2_A)$ and $V(q^2) = V(0)/(1-q^2/M^2_V)$, with pole masses $M_V=M_{D_s^*(1^-)}=2.1121$~GeV/$c^2$~\cite{pdg24} and $M_A=M_{D_{s1}(1^+)}=2.4595$~GeV/$c^2$~\cite{pdg24}.
The form factor $A_1(q^2)$ is common to all three helicity amplitudes. Therefore, it is natural to define two form factor ratios as $r_V=V(0)/A_1(0)$ and $r_2=A_2(0)/A_1(0)$
at the momentum transfer square $q^2=0$.

The amplitude of the $\mathcal{P}$-wave resonance $\mathcal{A}(m)$ is expressed as~\cite{prd94_032001,prd83_072001}
\begin{equation}
\mathcal{A}(m)=\frac{M_{K^{*}(892)}\Gamma^0_{K^{*}(892)}(p^*/p^*_0)}{m_{K^{*}(892)}^2-m^2_{\bar{K}^{0}\pi^-}-iM_{K^{*}(892)}\Gamma(m_{\bar{K}^{0}\pi^-})}\frac{B(p^*)}{B(p^*_0)},
\label{eq:bw}
\end{equation}
where $M_{K^{*}(892)}$ and $\Gamma^0_{K^{*}(892)}$ are the pole mass and decay width of the $K^{*}(892)^-$ resonance, respectively.
The Blatt-Weisskopf damping factor takes the form $B(p)=\frac{1}{\sqrt{1+R^2p^2}}$ with $R=3.07$~GeV$^{-1}$~\cite{prd94_032001}, and
$\Gamma \left(m_{\bar{K}^{0}\pi^-}\right)=\Gamma^0_{K^{*}(892)}\left(\frac{p^*}{p^*_0}\right)^3\frac{M_{K^{*}(892)}}{m_{\bar{K}^{0}\pi^-}}\left[\frac{B\left(p^*\right)}{B\left(p^*_0\right)}\right]^2$,
where $p^*_0$ is the momentum of the $\bar{K}^0$ at the pole mass of the $K^{*}(892)^-$ resonance. The parameter $\alpha$ given in Eq.~(\ref{eq:F2}) is defined by $\alpha=\sqrt{3\pi \mathcal{B}_{K^*}/(p^*_0\Gamma^0_{K^{*}(892)})}$, $\mathcal{B}_{K^*}=\mathcal{B}(K^{*}(892)^-\rightarrow \bar{K}^0\pi^-)=2/3$.

The $\mathcal{S}$-wave form-factor $\mathcal{F}_{10}$ is described by~\cite{prd94_032001,prd83_072001}
\begin{equation}
  \mathcal{F}_{10} = \left( p_{\bar{K}^{0}\pi^-}m_{D^0} \right)
    \left( \frac{1}{1-\frac{q^2}{m^2_A}} \right) \mathcal{A}_S(m),
\end{equation}
where $\mathcal{A}_S(m)$ corresponds to the mass-dependent $S$-wave amplitude. The expression $\mathcal{A}_S(m)=r_SP(m)e^{i\delta_S(m)}$ from Refs.~\cite{prd94_032001,prd83_072001} is adopted, in which $P(m)=1+xr_S^{(1)}$ with $x=\sqrt{\left(\frac{m}{m_{\bar{K}^0}+m_{\pi^-}}\right)^2-1}$, where $r_S$ and $r_S^{(1)}$ are the relative intensity and the dimensionless coefficient, respectively.  
The $\mathcal{S}$-wave phase $\delta_S(m)=\delta^{1/2}_{\rm BG}$ with $\cot(\delta^{1/2}_{\rm BG})=1/(a^{1/2}_{\rm S,BG}p^*)+b^{1/2}_{\rm S,BG}p^*/2$, where $a^{1/2}_{\rm S,BG}$ and $b^{1/2}_{\rm S,BG}$ are the scattering length and the effective range, respectively.

%%%%%%%%%%%%%%%%%%%%%%%%%%%%%
\begin{figure}[tp!]
\begin{center}
   \includegraphics[width=\linewidth]{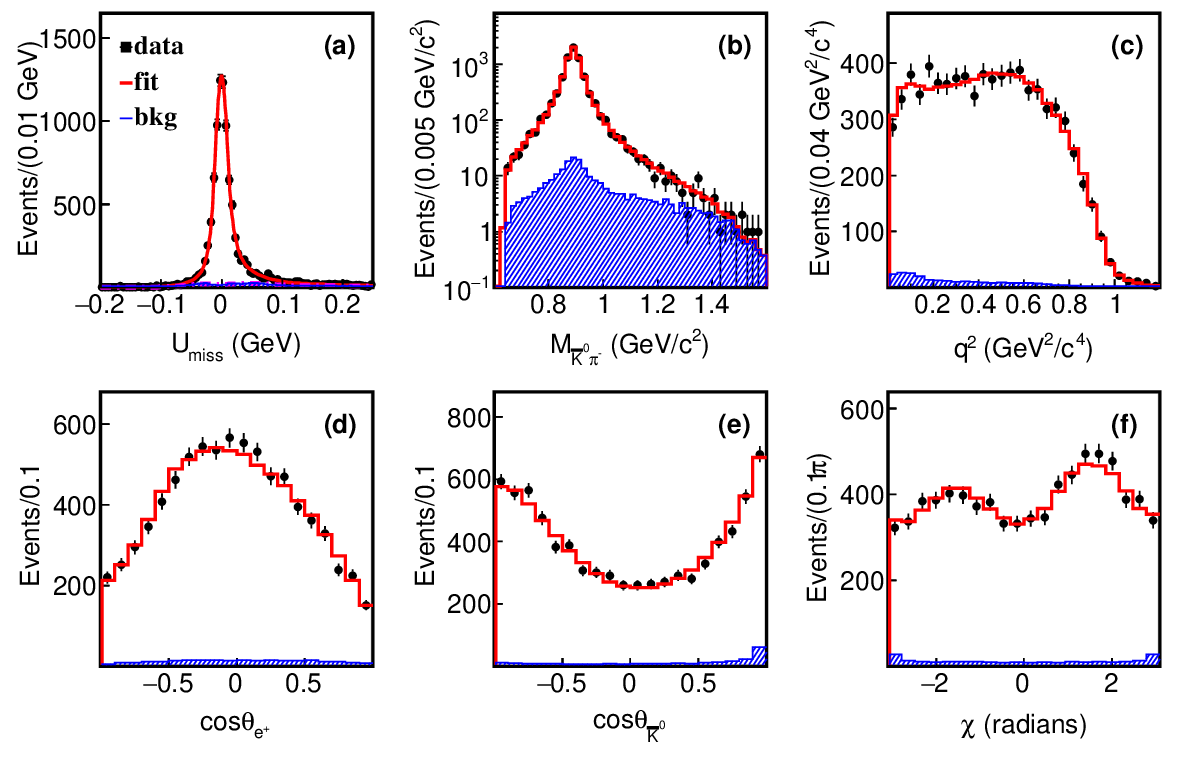}
   \caption{ (Color online)~(a) Fit to $U_{\rm miss}$ distribution of the SL candidate events.
   Distributions of the five kinematic variables (b) $M_{\bar{K}^0\pi^-}$, (c) $q^2$, (d) $\cos\theta_{e^+}$, (e) $\cos\theta_{\bar{K}^0}$, and (f) $\chi$ for SL decay $D^0\rightarrow \bar{K}^0\pi^-e^+\nu_e$. The dots with error bars are data, the red curves and histograms are the fit results, and the shaded histograms are the simulated background.}
\label{fig:formfactor}
\end{center}
\end{figure}
%%%%%%%%%%%%%%%%%%%%%%%%%%%%%

%%%%%%%%%%%%%%%%%%%%%%%%%%%%%
\begin{table}
\begin{center}
\caption{The fit results, where the first uncertainties are statistical and the second are systematic. } \normalsize
\begin{tabular}
{lc} \hline\hline  Variable~~~~~~~~~~~~~~~~~~~~~~~     &   Value    \\ \hline
$M_{K^{*}(892)^-}$ (MeV/$c^2$)                                      &   $892.3\pm0.5\pm0.2$          \\
$\Gamma_{K^{*}(892)^-}$ (MeV)                                      &  ~$46.5\pm0.8\pm0.2$          \\
$r_S$ (GeV)$^{-1}$                                                          & $-13.36\pm0.93\pm0.49$       \\
$a^{1/2}_{\rm S,BG}$ (GeV/$c$)$^{-1}$                          & ~~~$3.31\pm0.23\pm0.34$     \\
$r_S^{(1)}$                                                       &  ~$-0.04\pm0.06\pm0.03$       \\
$r_V$                                                               &  ~~~$1.48\pm0.05\pm0.02$     \\
$r_2$                                                               & ~~~$0.70\pm0.04\pm0.02$     \\
\hline\hline 
\end{tabular}
\label{tab:FitResults}
\end{center}
\end{table}
%%%%%%%%%%%%%%%%%%%%%%%%%%%%%

An unbinned five-dimensional maximum likelihood fit to the distributions of $m_{\bar{K}^0\pi^-}$, $q^2$, $\cos\theta_{e^+}$, $\cos\theta_{\bar{K}^0}$, and $\chi$ for the
$D^0\rightarrow \bar{K}^{0}\pi^- e^+\nu_{e}$ events within $-0.05<U_{\rm miss}<0.05$~GeV is performed in a similar manner to Ref.~\cite{prd94_032001,prd99_011103}.
The projected distributions of the fit onto the fitted variables are shown in figures~\ref{fig:formfactor}\,(b-f).
In this fit, the parameters $r_V$, $r_2$, $m_0$, $\Gamma_0$, $r_S$, $a^{1/2}_{\rm S,BG}$ and $r_S^{(1)}$ are free,
while $b^{1/2}_{\rm S,BG}$ is fixed to $-0.81$~(GeV/$c$)$^{-1}$, based on the analysis of $D^+\rightarrow K^+\pi^-e^+\nu_e$ at BESIII~\cite{prd94_032001}.
The fit results are summarized in table~\ref{tab:FitResults}.
The goodness of fit is estimated by using the $\chi^2/{\rm ndof}$, where ${\rm ndof}$ denotes the number of degrees of freedom. The $\chi^2$ is
calculated from the comparison between the measured and expected number of events in the five-dimensional space of the kinematic variables $m_{\bar{K}^0\pi^-}$, $q^2$, $\cos\theta_{e^{+}}$, $\cos\theta_{\bar{K}^0}$, and $\chi$ which are  divided into 2, 2, 3, 3, and 3 bins, respectively.
The bins are set with different sizes, so that they contain sufficient numbers of signal events for credible $\chi^2$ calculation. The $\chi^2$ value is
calculated as
\begin{equation}
\chi^2=\displaystyle{\sum_i^{\rm N_{\rm bin}}\frac{(n_i^{\rm data}-n_i^{\rm fit})^2}{n_i^{\rm data}}},
\end{equation}
where $N_{\rm bin}$ is the number of bins, $n_i^{\rm data}$ denotes the
measured number of events of the $i$-th bin, and $n_i^{\rm fit}$ denotes the expected number of events of the $i$th bin. The ${\rm ndof}$ is defined as the number of bins minus
the number of fit parameters. The $\chi^2/{\rm ndof}$ obtained is 113.7/101. The fit procedure is validated using a large simulated sample of inclusive events, where the pull distribution of each fitted parameter is found to be consistent with a normal distribution.

The fit fraction of each component can be determined by the ratio of the decay intensity of the specific component to that of the total intensity. The fractions of $\mathcal{S}$-wave and $\mathcal{P}$-wave ($K^{*}(892)^-$) are 
$f_{S-{\rm wave}}=(5.87\pm0.32({\rm stat}))\%$ and $f_{K^{*}(892)^-}=(94.15\pm0.32({\rm stat}))\%$, respectively, where the uncertainty propagation includes correlations among the underlying parameters.  

%%%%%%%%%%%%%%%%%%%%%%%%%%%%%
\begin{table*}
\begin{center}
\caption{Systematic uncertainties on the fitted parameters, in \%. }
\resizebox{!}{2.1cm}{
\begin{tabular}
{lcccccccccc} \hline\hline \normalsize
Parameter~~~~~& ~$E_{\gamma\,{\rm \max}}$~ &   ~$M_{\bar{K}^0\pi^-e^+}$~  & ~~$E/p$~~ & ~~~~$f$~~~~ & ~${\rm Tracking, PID}$~ & ~$D$-wave~ & ~$b^{1/2}_{\rm S,BG}$~ & ~~${\rm Total}$~~ \\ \hline
$M_{K^{*}(892)^-}$            &   0.01 & 0.01 & 0.01 & 0.00 & 0.01 & 0.01 & 0.01 &~~0.02  \\
$\Gamma_{K^{*}(892)^-}$ &   0.36 & 0.17 & 0.30 & 0.11 & 0.37 & 0.15 & 0.44 &~~0.78  \\
$r_S$                                 &   0.47 & 1.16 & 2.37 & 1.30 & 1.26 & 1.74 & 0.01 &~~3.67  \\
$a^{1/2}_{\rm S,BG}$        &   0.12 & 0.95 & 0.27 & 0.29 & 0.28 & 3.01 & 10.0 & ~~10.5  \\
$r^{(1)}_S$                        &  17.4  & 26.8 & 46.0 & 43.4 & 18.6 & 17.1 & 17.7 & ~~77.3  \\
$r_V$                                 &   0.17 & 0.97 & 0.66 & 0.14 & 0.69 & 0.14 & 0.22 &~~1.40  \\
$r_2$                                 &   0.70 & 0.95 & 1.92 & 0.21 & 0.07 & 0.01 & 2.59 &~~3.44  \\
$f_{K^{*}(892)^-}$             &   0.05 & 0.04 & 0.01 & 0.08 & 0.07 & 0.08 & 0.06 &  ~~0.17  \\
$f_{S-{\rm wave}}$            &  0.85  & 0.68 & 0.17 & 1.36 & 1.19 & 1.36 & 0.85 & ~~2.66  \\
\hline\hline
\end{tabular}
}
\label{tab:Syserr}
\end{center}
\end{table*}
%%%%%%%%%%%%%%%%%%%%%%%%%%%%%

The systematic uncertainties of the fitted parameters and the fractions of $S$-wave and $K^{*}(892)^-$ components
are defined as the difference between the fit results in nominal conditions and those obtained with varied conditions.
The systematic uncertainties due to the $E_{\gamma\,{\rm \max}}$, $M_{\bar{K}^0\pi^-e^+}$ and $E/p$ requirements are estimated by using alternative requirements of
$E_{\gamma\,{\rm \max}}<0.20$~GeV, $M_{\bar{K}^0\pi^-e^+}<1.75$~GeV/$c^2$ and $E/p>0.75$, respectively. 
The systematic uncertainty due to the background fraction is estimated by varying its value by $\pm 10\%$
which accounts for the possible difference of the background fraction in the selected $U_{\rm miss}$ region.
The systematic uncertainties arising from the tracking and PID placed on the charged pion, the electron and the $K^0_S$ are estimated by varying the pion and electron tracking and PID efficiencies, and 
$K^0_S$ detection efficiency by $\pm0.5\%$, $\pm0.5\%$ and $\pm0.6\%$, respectively.
The systematic uncertainty due to neglecting a possible contribution from the $D$-wave component is estimated by incorporating the $D$-wave component in Eq.~(\ref{eq:F1}). The systematic uncertainty in the fixed parameter $b^{1/2}_{\rm S,BG}$ is estimated
by varying its nominal values by $\pm1\sigma$. 
All of the variations mentioned above will result in differences of the fitted parameters and the extracted fractions of $\mathcal{S}$-wave and $K^{*}(892)^-$ components from that under the nominal conditions.  These differences are assigned as the systematic uncertainties and 
summarized in table~\ref{tab:Syserr}, where the total systematic uncertainty is obtained by adding all contributions in quadrature.

%%%%%%%%%%%%%%%%%%%%%%%%%%%%%%%%%%%%%%%%%%%%%%%%%%%%%%%%%%%%%%%%
%%%%%%%%%%%%%    summary       Part                %%%%%%%%%%%%%
%%%%%%%%%%%%%%%%%%%%%%%%%%%%%%%%%%%%%%%%%%%%%%%%%%%%%%%%%%%%%%%%
\section{Summary}
\begin{figure}[tp!]
\begin{center}
   %\flushleft
   \begin{minipage}[t]{7cm}
   \includegraphics[width=\linewidth]{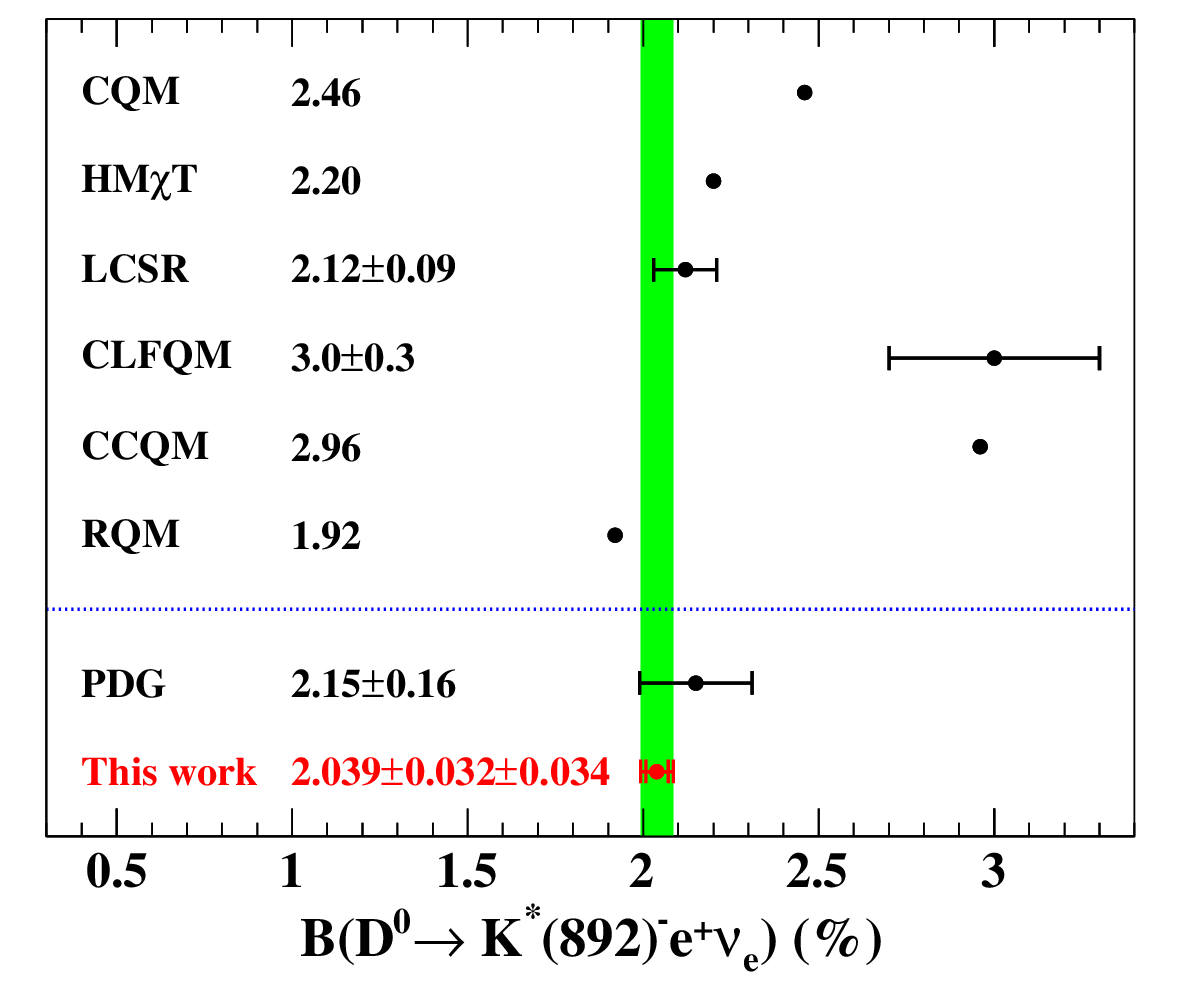}
   \put(-28,150){\bf  (a)}
   \end{minipage}
   \begin{minipage}[t]{7cm}
   \includegraphics[width=\linewidth]{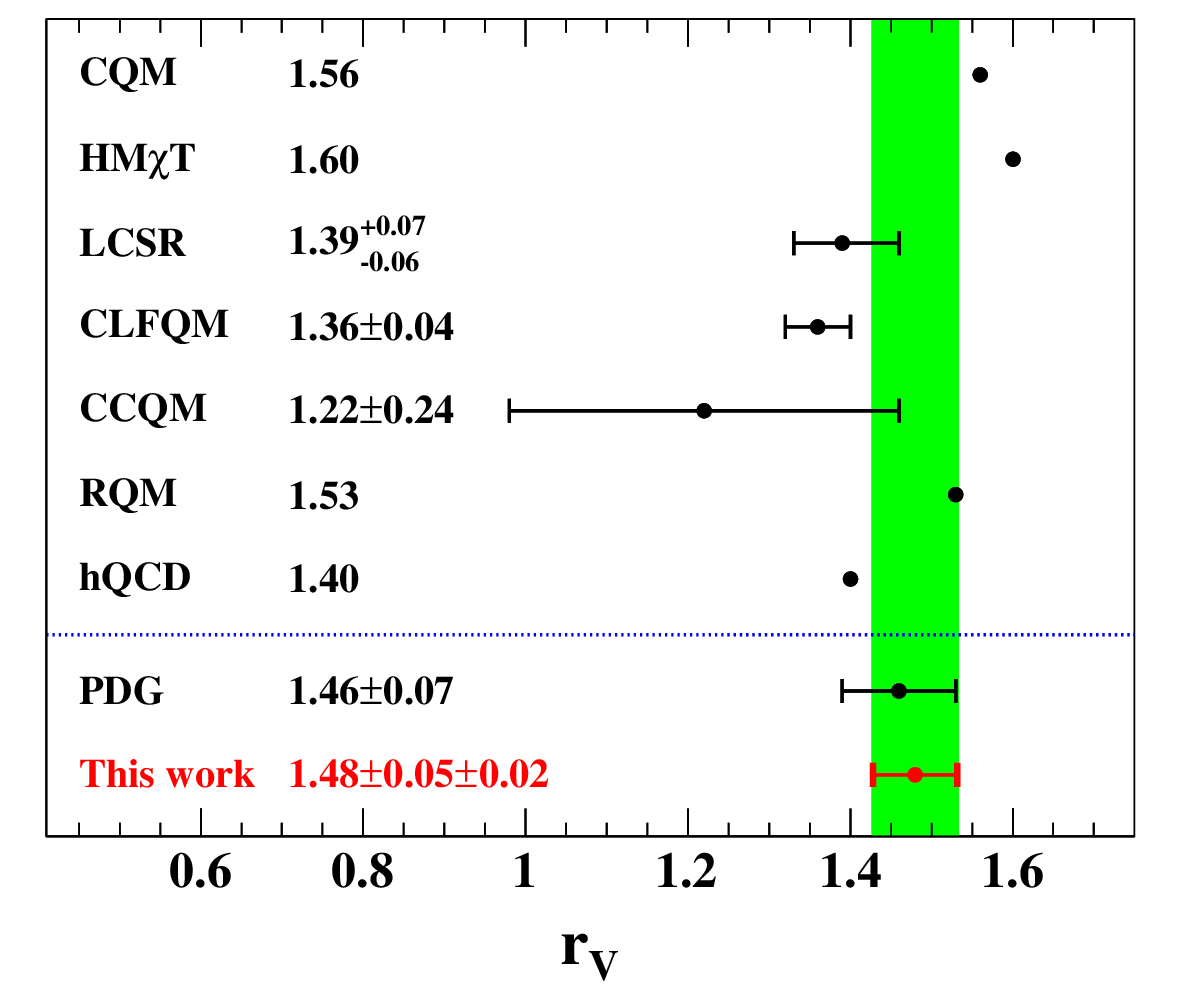}
    \put(-28,150){\bf  (b)}
   \end{minipage}   
   \begin{minipage}[t]{7cm}
   \includegraphics[width=\linewidth]{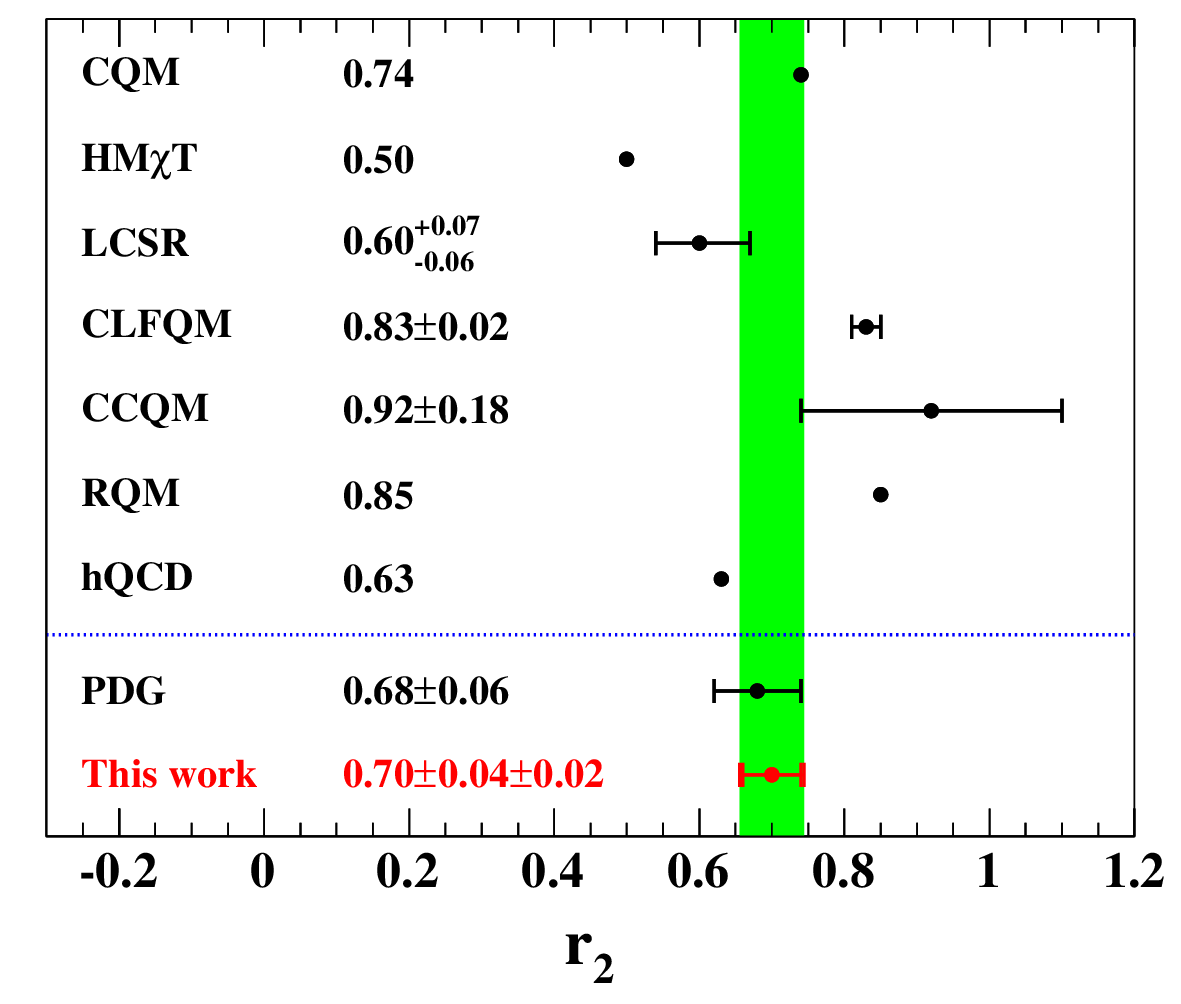}
    \put(-28,150){\bf  (c)}
   \end{minipage}   
   \begin{minipage}[t]{7cm}
   \includegraphics[width=\linewidth]{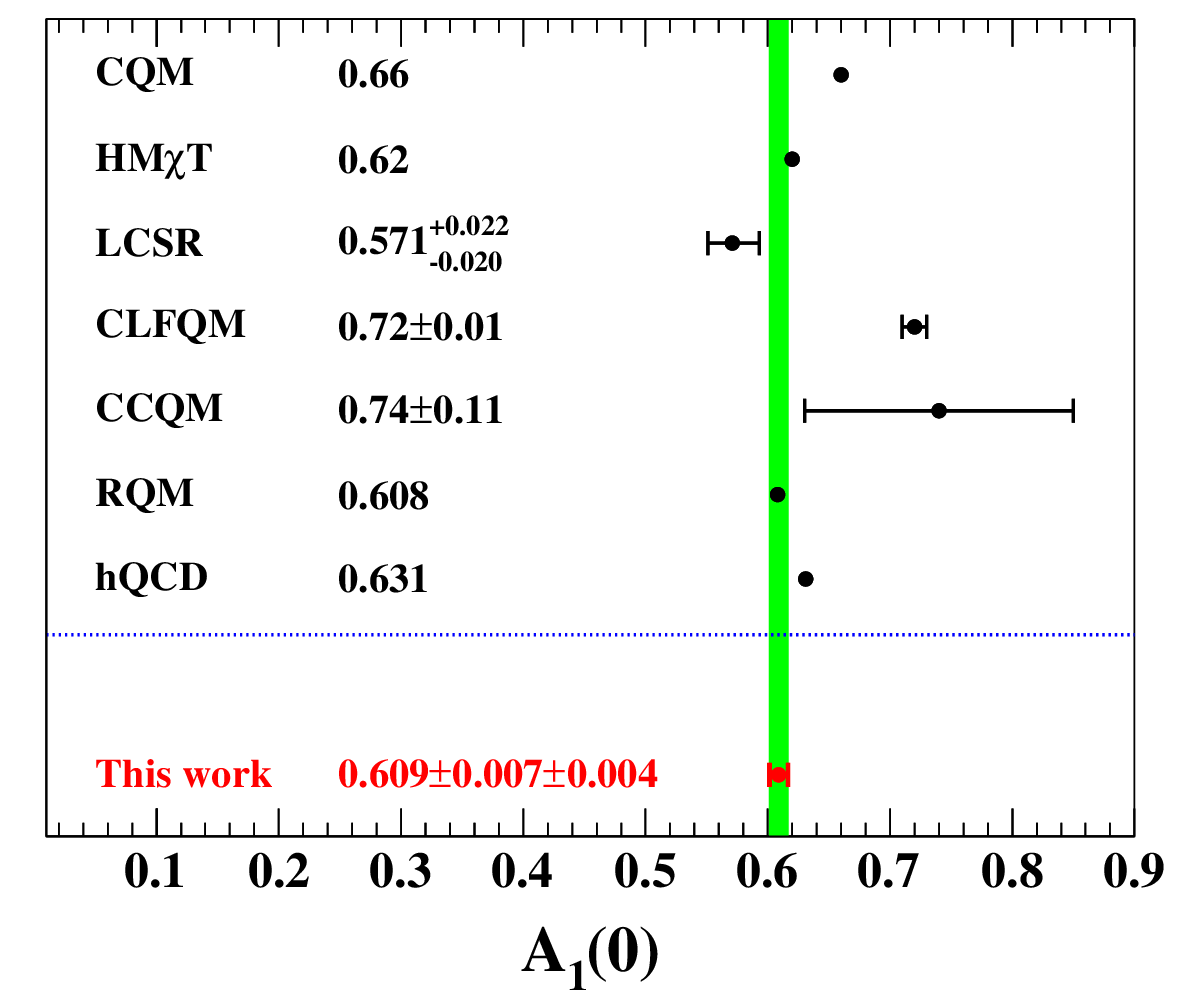}
   \put(-28,150){\bf  (d)}
   \end{minipage}         
   \caption{ (Color online)~Comparisons of the measured (a) $\mathcal{B}(D^0\rightarrow K^{*}(892)^-e^+\nu_e)$, (b) $r_V$, (c) $r_2$ and (d) $A_1(0)$ with various theoretical calculations from CQM~\cite{PRD62_014006}, HM$\chi$T~\cite{PRD72_034029}, LCSR~\cite{IJMPA21_6125}, CLFQM~\cite{JPG39_025005,EPJC77_587}, CCQM~\cite{FrontPhys14_64401}, RQM~\cite{prd101_013004}, hQCD~\cite{prd109_026008} and PDG averaged results~\cite{pdg24}. }
\label{fig:cmpformfactor}
\end{center}
\end{figure}

In summary, using $7.9~\mathrm{fb}^{-1}$ of $e^+e^-$ annihilation data collected at $\sqrt{s}=3.773$ GeV by the BESIII detector, the absolute BF of $D^0\rightarrow \bar{K}^0\pi^-e^+\nu_{e}$ is measured to be $\mathcal{B}(D^0\rightarrow \bar{K}^0\pi^-e^+\nu_{e})=  (1.444 \pm 0.022_{\rm stat} \pm 0.024_{\rm syst})\%$, which presents the most precise measurements to date~\cite{pdg24}.
By analyzing the dynamics of $D^0\rightarrow \bar{K}^0\pi^-e^+\nu_{e}$ decay, the $\mathcal{S}$-wave component is measured with a fraction $f_{S-{\rm wave}} = (5.87 \pm 0.32_{\rm stat} \pm 0.16_{\rm syst})\%$, leading to $\mathcal{B}[D^0\rightarrow (\bar{K}^0\pi^-)_{S-{\rm wave}}e^+\nu_e] = (0.085 \pm 0.005_{\rm stat} \pm 0.003_{\rm syst})\%$.
The $\mathcal{P}$-wave component is observed with a fraction of $f_{K^{*}(892)^-}=(94.15 \pm 0.32_{\rm stat} \pm 0.16_{\rm syst})\%$ and the corresponding BF is given as $\mathcal{B}(D^0\rightarrow K^*(892)^-e^+\nu_e) = (2.039 \pm 0.032_{\rm stat} \pm 0.034_{\rm syst})\%$ with $\mathcal{B}(K^{*}(892)^-\rightarrow \bar{K}^0\pi^-)=2/3$.
In addition, the form factor ratios of the $D^0\rightarrow K^{*}(892)^-e^+\nu_{e}$ decay are determined to be
$r_V = 1.48 \pm 0.05_{\rm stat} \pm 0.02_{\rm syst}$ and
$r_2 = 0.70 \pm 0.04_{\rm stat} \pm 0.02_{\rm syst}$.
The reported results in this work are consistent with, and more precise than the previous measurements and hence supersede results reported in Ref.~\cite{prd99_011103}.  
The comparisons of the measured BF and form-factor parameters between this measurement and theoretical calculations are shown in table~\ref{tab:cmpresult} and figure~\ref{fig:cmpformfactor}.
At a confidence level of 95\%, our measured BF disfavors the central values in Refs.~\cite{PRD62_014006,FrontPhys14_64401,EPJC77_587}, and the measured $r_2$ disfavors the central values in Refs.~\cite{PRD72_034029,JPG39_025005,prd101_013004}.
Note that unlike Ref.~\cite{prd99_011103}, the parameter $r_S^{(1)}$ is now free in the fit, giving a more general parameterization for the $\mathcal{S}-$wave component in the present work.  This affects the comparison of  uncertainties on the $r_V$ and $r_2$ parameters in the table~\ref{tab:cmpresult}, where the PDG results are dominated by Ref.~\cite{prd99_011103}.  

The value of $A_1(0)$ can be obtained by integrating Eq.~(\ref{eq:differential}), restricted to $K^*(892)^{-}$ contribution, over the three angles:
\begin{equation}
\frac{d^2\Gamma}{dq^2dm^2}=\frac{2}{9}\frac{G^2_F|V_{cs}|^2}{(4\pi)^5m^2_{D^0}}p_{\bar{K}^0\pi}\beta (|\mathcal{F}_{11}|^2+|\mathcal{F}_{21}|^2+|\mathcal{F}_{31}|^2 ).
\label{eq:inte}
\end{equation}
The decay width of $D^0\rightarrow K^{*}(892)^-e^+\nu_{e}$ is related to $\mathcal{B}(D^0\rightarrow K^{*}(892)^-e^+\nu_{e})$ measured in this work and the lifetime by
\begin{equation}
\Gamma=\frac{\hbar\mathcal{B}(D^0\rightarrow K^{*}(892)^-e^+\nu_{e})\mathcal{B}(K^{*}(892)^-\rightarrow \bar{K}^0\pi^-)}{\tau_{D^0}}=|A_1(0)|^2|V_{cs}|^2\times \mathbb{X}.
\label{eq:gama}
\end{equation}
Here $\mathbb{X}=\frac{16}{9}\frac{G^2_F}{(4\pi)^5m^2_{D^0}}\int_0^{q^2_{\rm max}}\int_{m_{\rm min}^2}^{m^2_{\rm max}}  p_{\bar{K}^0\pi}\beta \alpha^2q^2 \times |\mathcal{A}(m)|^2\frac{|H_0|^2+|H_+|^2+|H_-|^2}{|A_1(0)|^2}dm^2dq^2$,
where $q^2_{\max}=(m_{D^0}-m_{\bar{K}^0}-m_{\pi})^2$, and $m_{\bar{K}^0}$ and $m_{\pi}$ are the known $\bar{K}^0$ and $\pi^-$ masses~\cite{pdg24}. 
The integration region over $m^2$ ranges from $m^2_{\rm min}=(m_{\bar{K}^0}+m_{\pi})^2$ to $m^2_{\rm max}=1.80^2$~GeV$^2/c^4$.
Then the integration in Eq.~(\ref{eq:gama}) can be calculated using the form factor parameters measured in this work. Using the $\tau_{D^0}=410.3\pm1.0~{\rm fs}$ and $|V_{cs}|=0.97435\pm0.00016$~\cite{pdg24}, we obtain $A_1(0)=0.610\pm0.007\pm0.004$, which is the first determination using the data of $D^0\rightarrow K^{*}(892)^-e^+\nu_{e}$ decay.
The comparisons of the $A_1(0)$ between this measurement and theoretical calculations are also shown in table~\ref{tab:cmpresult} and figure~\ref{fig:cmpformfactor}.

\begin{table}
\begin{center}
\caption{Comparisons of the measured BF and form-factor parameters for $D^0\rightarrow K^{*}(892)^-e^+\nu_e$, with various theoretical calculations and PDG averaged results. }
\resizebox{!}{2.0cm}{
\begin{tabular} {|l|c|c|c|c|} \hline  \tiny
& $\mathcal{B}_{D^0\rightarrow K^{*}(892)^-e^+\nu_e}$ (\%)  & $r_V$  & $r_2$ & $A_1(0)$  \\ \hline
This work                                    & $2.039\pm0.032\pm0.034$ & $1.48\pm0.05\pm0.02$ & $0.70\pm0.04\pm0.02$ & $0.610\pm0.007\pm0.004$ \\
CQM~\cite{PRD62_014006}       & 2.46                     & 1.56 & 0.74 & 0.66 \\
HM$\chi$T~\cite{PRD72_034029} & 2.20                       & 1.60 & 0.50 & 0.62 \\
LCSR~\cite{IJMPA21_6125}         &   $2.12\pm0.09$  & $1.39^{+0.06}_{-0.07}$  & $0.60^{+0.06}_{-0.07}$ &$0.571^{+0.020}_{-0.022}$  \\
CLFQM~\cite{JPG39_025005,EPJC77_587}     &   $3.0\pm0.3$      & $1.36\pm0.04$ & $0.83\pm0.02$ & $0.72\pm0.01$ \\ 
CCQM~\cite{FrontPhys14_64401} &  2.96                    & $1.22\pm0.24$ & $0.92\pm0.18$ & $0.74\pm0.11$ \\
RQM~\cite{prd101_013004}        &  1.92                     & 1.53 & 0.85 & 0.608 \\ 
hQCD~\cite{prd109_026008}       & $--$                      &  1.40 & 0.63 & 0.631  \\
PDG 2024~\cite{pdg24}             &   $2.15\pm0.16$  & $1.46\pm0.07$ & $0.68\pm0.06$ & $--$ \\
\hline
\end{tabular}
}
\label{tab:cmpresult}
\end{center}
\end{table}

%%%%%%%%%%%%%%%%%%%%%%%%%%%%%
%%%%%%%%%%%%%%%%%%%%%%%%%%%%%

%%%%%%%%%%%%%%%%%%%%%%%%%%%%%%%%%%%%%%%%%%%%%%%%%%%%%%%%%%%%%%%%
%%%%%    acknowledgments       Part                %%%%%%%%%%%%%
%%%%%%%%%%%%%%%%%%%%%%%%%%%%%%%%%%%%%%%%%%%%%%%%%%%%%%%%%%%%%%%%
\acknowledgments
The BESIII Collaboration thanks the staff of BEPCII and the IHEP computing center for their strong support. This work is supported in part by National Key R\&D Program of China under Contracts Nos. 2023YFA1606000, 2020YFA0406300, 2020YFA0406400; National Natural Science Foundation of China (NSFC) under Contracts Nos. 11635010, 11735014, 11935015, 11935016, 11935018, 12022510, 12025502, 12035009, 12035013, 12061131003, 12192260, 12192261, 12192262, 12192263, 12192264, 12192265, 12221005, 12225509, 12235017, 12375090, 12361141819; the Chinese Academy of Sciences (CAS) Large-Scale Scientific Facility Program; the CAS Center for Excellence in Particle Physics (CCEPP); Joint Large-Scale Scientific Facility Funds of the NSFC and CAS under Contract No. U1832207; 100 Talents Program of CAS; The Institute of Nuclear and Particle Physics (INPAC) and Shanghai Key Laboratory for Particle Physics and Cosmology; German Research Foundation DFG under Contracts Nos. FOR5327, GRK 2149; Istituto Nazionale di Fisica Nucleare, Italy; Knut and Alice Wallenberg Foundation under Contracts Nos. 2021.0174, 2021.0299; Ministry of Development of Turkey under Contract No. DPT2006K-120470; National Research Foundation of Korea under Contract No. NRF-2022R1A2C1092335; National Science and Technology fund of Mongolia; National Science Research and Innovation Fund (NSRF) via the Program Management Unit for Human Resources \& Institutional Development, Research and Innovation of Thailand under Contracts Nos. B16F640076, B50G670107; Polish National Science Centre under Contract No. 2019/35/O/ST2/02907; Swedish Research Council under Contract No. 2019.04595; The Swedish Foundation for International Cooperation in Research and Higher Education under Contract No. CH2018-7756; U. S. Department of Energy under Contract No. DE-FG02-05ER41374

%%%%%%%%%%%%%%%%%%%%%%%%%%%%%%%%%%%%%%%%%%%%%%%%%%%%%%%%%%%%%%%%
%%%%%    bibliographies       Part                %%%%%%%%%%%%%
%%%%%%%%%%%%%%%%%%%%%%%%%%%%%%%%%%%%%%%%%%%%%%%%%%%%%%%%%%%%%%%%

\newpage
\input{authorlist_2024-08-01.tex}
%\end{linenumbers}
\end{document}

%% file: authorlist_2024-08-01.tex
%% Saved at => 2024-08-01
M.~Ablikim$^{1}$, M.~N.~Achasov$^{4,c}$, P.~Adlarson$^{76}$, O.~Afedulidis$^{3}$, X.~C.~Ai$^{81}$, R.~Aliberti$^{35}$, A.~Amoroso$^{75A,75C}$, Y.~Bai$^{57}$, O.~Bakina$^{36}$, I.~Balossino$^{29A}$
% ...
\\
\vspace{0.2cm}
(BESIII Collaboration)\\
\vspace{0.2cm} {\it
$^{1}$ Institute of High Energy Physics, Beijing 100049, People's Republic of China\\
% ...
$^{3}$ Bochum  Ruhr-University, D-44780 Bochum, Germany\\
$^{4}$ Budker Institute of Nuclear Physics SB RAS (BINP), Novosibirsk 630090, Russia\\
% ...
$^{29}$ INFN Sezione di Ferrara, (A)INFN Sezione di Ferrara, I-44122, Ferrara, Italy; (B)University of Ferrara,  I-44122, Ferrara, Italy\\
% ...
$^{35}$ Johannes Gutenberg University of Mainz, Johann-Joachim-Becher-Weg 45, D-55099 Mainz, Germany\\
$^{36}$ Joint Institute for Nuclear Research, 141980 Dubna, Moscow region, Russia\\
% ...
$^{57}$ Southeast University, Nanjing 211100, People's Republic of China\\
% ...
$^{75}$ University of Turin and INFN, (A)University of Turin, I-10125, Turin, Italy; (B)University of Eastern Piedmont, I-15121, Alessandria, Italy; (C)INFN, I-10125, Turin, Italy\\
$^{76}$ Uppsala University, Box 516, SE-75120 Uppsala, Sweden\\
% ...
$^{81}$ Zhengzhou University, Zhengzhou 450001, People's Republic of China\\

\vspace{0.2cm}
% ...
$^{c}$ Also at the Novosibirsk State University, Novosibirsk, 630090, Russia\\
% ...

}
%% ends here %%

%% file: arXiv_D0Kspiev_20241214.bbl
\begin{thebibliography}{99}
\bibitem{chargeneutral} Throughout this paper, mesons with no explicit charge denote either a charged or neutral meson;
the charge-conjugate modes are implied, unless explicitly noted.
\bibitem{BFK0} The BFs of $K^0\rightarrow K^0_S$ and $K^0_S\rightarrow \pi^+\pi^-$ are included in the detection efficiency $\epsilon_{\rm DT}$.  The $K^0_S$ fraction of a $K^0$ is taken as 50\% and the BF of $K^0_S\rightarrow \pi^+\pi^-$ is taken from PDG~\cite{pdg24}.
\end{thebibliography}
